# A Priority-Based Cross-Layer Design for Future VANETs through Full-Duplex Technology

Junwei Zang*, Vahid Towhidlou and Mohammad Shikh-Bahaei

*Abstract*—Among all requirements for vehicle-to-everything (V2X) communications, successful delivery of packets with small delay is of the highest significance. Especially, the delivery of a message before a potential accident (i.e. emergency message) should be guaranteed. In this work, we propose a novel cross-layer design to enhance the delivery of emergency messages so that accidents can be further avoided. Particularly, in the PHY layer, imperfect full-duplex (FD) simultaneous transmitting and sensing is analysed and dynamic thresholds for determining the channel status before and during transmissions are mathematically formulated. Then a novel FD MAC protocol, named priority-based multiple access (PBMA), based on prioritised messaging between different vehicles is proposed. Average collision probability, collision duration, waiting time as well as successful delivery rate of the system are formulated too. The delivery performance of emergency messages is also mathematically derived. In addition, comparisons have been made among three different mechanisms. Benchmark is the DSRC standard which uses half-duplex (HD) technology with enhanced distributed channel access (EDCA) protocol. We also compare our proposed protocol with FD EDCA. Simulations have verified the accuracy of our analysis. They have also illustrated the delivery of emergency messages has been enhanced by deploying our proposed design.

*Index Terms*—Full Duplex, DSRC, V2X, VANETs, IEEE 802.11p, connected vehicles.

## I. INTRODUCTION

VEHICLE-TO-EVERYTHING (V2X) communication has been proposed to reduce accidents in future intelligent transportation system (ITS). Two promising standards are considered as potential candidates. One of them is IEEE 802.11p which is also known as the PHY/MAC specifications of dedicated short-range communication (DSRC) [1]; while the

This is an Arxiv version of the transaction paper. The final version will be available soon on IEEE Transactions on Vehicular Technology.

Corresponding author*

Junwei Zang, Vahid Towhidlou and Mohammad Shikh-Bahaei are with the Department of Engineering, Centre for Telecommunications Research in King's College London, London, U.K. (e-mail addresses: {junwei.zang, vahid.towhidlou, m.sbahaei}@kcl.ac.uk)

This work is partially supported by EPSRC SENSE Project under grant number EP/P003486/1.

other solution is cellular-V2X (C-V2X) [2], which is built upon 4G or 5G platforms. Comparison between these two standards has been extensively studied in many works such as [3]. However, the goal of having an unified method for future V2X networks is not accomplished yet due to the fact that each of them has its own advantages. This work is built upon DSRC.

In current DSRC standard, safety messages known as cooperative awareness messages (CAMs) are exchanged periodically between vehicles in a broadcasting manner. In this case, acknowledgement (ACK) messages cannot be used to detect a failed transmission, and ACK messaging is not incorporated in the DSRC standard. In other words, nodes cannot detect collisions, and loss of these CAMs due to concurrent broadcasting would lead to a higher risk of accident. Current enhanced distributed channel access (EDCA) method adopted by DSRC needs further improvement, since the performance is shown to degrade significantly when the number of nodes in the network increases (dense network) [4]-[5].

Another weak aspect of DSRC is related to the priority of CAMs. The levels of priority are named as access categories (ACs). Four ACs are defined (AC0-AC3), where AC0 has the highest priority and AC3 has the lowest priority. In addition, a backoff mechanism is used to avoid collisions. Smaller maximum contention window (CW) size and arbitrary inter-frame space (AIFS) are used to differentiate a high-priority AC from a low-priority AC. Therefore, high-priority ACs will access channel with a higher probability, and are expected to experience less backoff time and total waiting time. This collision avoidance and prioritised messaging mechanism adopted by EDCA is also known as internal collision handling mechanism [6]-[7], as shown in Fig. 1. [6]. However, if CAMs with different priority levels are not generated at the same time, EDCA even cannot provide a higher access probability to the vehicle which has a high-priority CAM. The vehicle which has a low-priority CAM may complete the backoff process and take the channel before the high-priority CAM. In addition, when a collision has already happened between CAMs from different vehicles with different priority levels (a.k.a. external collision [6]-[7]), all colliding CAMs are lost, because vehicles neither can recognise the collision, nor identify the priority of the colliding CAM. In other words, when multiple vehicles have selected the same slot to broadcast, EDCA neither can guarantee the transmission of high-priority CAMs, nor provide high-priority CAMs a greater probability to broadcast. The reason is that a vehicle is not able to detect concurrent transmissions, and it is also not able to identify the priority of a CAM from another vehicle. So emergency messages could be delayed or even lost, which would in turn result in a higher risk of accident.

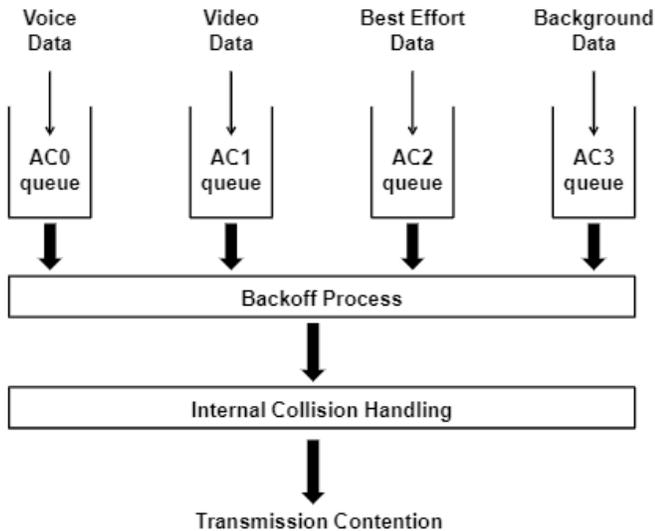

Fig. 1. EDCA prioritised channel access [6]

There has been extensive work on CAM broadcast in DSRC. H. Luong et al. focused on optimising the CAM broadcast repetition interval [8]. F. Lyu et al. proposed a time slot-sharing MAC protocol for CAM broadcast [9]. They defined a danger coefficient for investigating the rear-end collisions, and then proposed a distributed congestion control scheme in [10]. Afterwards, they have collected DSRC communication traces, and proposed a beaconing scheme in [11] to enhance broadcast reliability. In addition, Z. Tong et al. worked on the modeling of DSRC by using stochastic geometry [12]. Besides, many other works are showing potential congestion control mechanisms for future vehicular networks [13], [14]-[15]. Meanwhile, in the PHY layer, several control techniques can be considered for better spectral efficiency, such as [16]-[17]. In the MAC layer, dynamic optimisation of re-try limit in conjunction with link adaptation to achieve higher throughput at low frame loss rate can also considered for future VANETs [18]. However, none of these works applied in-band full-duplex (FD) technology [19] to V2X networks. Thus, vehicles cannot detect collisions whilst transmitting. An overview of FD technology in vehicular communications is provided in [20]. In addition, an example of FD simultaneous sensing and transmitting is provided in [21]. Furthermore, M. Yang et al. used FD technology in vehicular networks and focused on managing the interference. Their results have shown the feasibility of deploying FD technology in future V2X networks [22]. To the best of our knowledge, the most recent and relevant work is [23], in which A. Bazzi et al. introduced FD technology and an enhanced CSMA/CA protocol into vehicular communications. Their results have shown the effectiveness on the reliability of deploying FD technology in terms of collision probability of packets and packet delivery ratio. However, the operation of simultaneous transmission and sensing was considered to be perfect, which is not realistic. In addition, their proposed MAC layer protocol did not differentiate the priority of CAMs between vehicles, hence low-priority messages could still transmit before high-priority messages. In other words, emergency messages, which are supposed to be transmitted before normal update messages to prevent accidents, may be delayed.

In order to conquer the aforementioned challenges, we propose to equip on-board units (OBUs) with our design. By deploying FD technology, vehicles are able to broadcast CAMs and sense the channel status at the same time over the same frequency band. Sensing is carried out through measuring the energy level of the channel, which is the simplest and the most widely applied sensing method. Then vehicles can react to message collisions as soon as collisions are detected. Due to the fact that the existing prioritisation (i.e. ACs) in DSRC is designed for internal collisions, we introduce a new prioritisation scheme for detecting external collisions, and for enhancing the transmission of high-priority CAMs, when external collisions have happened. After detecting another ongoing transmission prior to a vehicle's own transmission, or collisions during a vehicle's transmission, proper actions should be followed immediately. We also propose a novel FD MAC protocol named priority-based multiple access (PBMA) to further schedule transmissions according to the priority of colliding CAMs. A vehicle which is in an emergency situation has an opportunity to re-attempt to broadcast immediately before going through a backoff process whilst periodic update CAMs defer their transmission normally according to the same backoff rules in DSRC.

The contributions of this work are summarised as follows.

- We extend the PHY layer sensing method in our recent work in [24] by analysing the Doppler effect. Furthermore, herein we introduce a novel cross-layer design across PHY and MAC layers for efficient vehicular communications. In particular, here we propose a novel prioritisation scheme which is dedicated to CAMs and a MAC layer protocol named PBMA.
- Unlike other works such as [23] and [25], FD sensing results are not assumed to be perfect. Dynamic thresholds and increased sensing window size for deciding the channel status before and during broadcasting are mathematically derived. Closed-form expressions of detection and false alarm probabilities are also found based on transmit power, target probabilities of detection and false alarm, sensing time and self-interference cancellation (SIC) capability.
- Based on our design, a vehicle can enjoy prioritised CAM messaging when competing with other vehicles' broadcast messages. When a collision is detected, emergency messages can re-attempt to broadcast immediately before going to the backoff process. Average collision probability, collision duration, waiting time as well as successful delivery rate of the system are formulated. The delivery of emergency messages is also analysed through mathematics and simulations, and its performance is shown to be enhanced.

The rest of the paper is organised as follows. Section II explains our proposed prioritisation scheme and the PBMA protocol. Section III describes the system model including assumptions and important notations. Corresponding mathematical analysis of both PHY and MAC layers are given in Section IV. Based on the mathematical analysis, numerical simulations are conducted and discussed in Section V. Finally, the paper is concluded in Section VI.

## II. THE PROPOSED PBMA PROTOCOL

In order to cope with the priority issue between messages from different vehicles, we propose a novel MAC protocol named priority-based multiple access (PBMA) mechanism, in which FD technology is used for simultaneous transmission and sensing. First of all, a new prioritisation of CAMs is proposed for external collision detection and external transmission contention. We categorise CAMs into three types: critical CAMs (*Mc*), emergency CAMs (*Me*) and normal vehicle status update CAMs (*Mn*). Mc CAMs have the highest priority and *Mn* CAMs have the lowest priority. The relationship between them and example generation scenarios of each type of CAMs are detailed in Table. I. All types of CAMs contain information about the status of the vehicle such as speed and location. However, any sudden change of the status of the vehicle, such as a harsh breaking, is a critical and dangerous activity which should be sent out as soon as possible (generation scenario of *Mc*). Emergency messages (*Me*) are generated when other gentle manoeuvres happen such as lane merging, because this type of message also shows the change of the vehicle status from normal cruising activity, which should also be recognised by other vehicles as soon as possible. The third category of CAMs (*Mn*) is generated when there is no potential danger and the vehicle is moving smoothly. In summary, we prioritise CAMs according to the status of the vehicle. Critical CAMs and emergency CAMs are event-triggered, vehicles would generate normal update CAMs for most of the time. Our proposed categorisation only shows one approach to differentiate priorities between vehicles, any other number of priority levels could be considered.

TABLE I
PRIORITISATION OF SAFETY MESSAGES

| Priority | CAM type | Example Generation Scenario |
|---|---|---|
| High | *Mc* | Harsh breaking, skidding etc. |
| Middle | *Me* | lane merging, gentle breaking etc. |
| Low | *Mn* | normal cruising etc. |

The proposed PBMA protocol operates as follows. When a CAM is generated at a generic vehicle, same as in legacy HD EDCA mechanism, it first probes the medium for $t_{sens}$ to determine whether the channel is busy or idle. If the channel is idle, the CAM is broadcast immediately in a FD manner, i.e. transmit and sense at the same time (TS mode). Otherwise, if the channel is occupied, normal update messages would defer its transmission by a random time interval (the backoff process) same as in the accessing rule in DSRC; critical and emergency messages would keep sensing the channel instead of initiating the backoff process, and start broadcasting as soon as the channel is sensed idle for $t_{sens}$ seconds.

During the transmission, unlike HD EDCA mechanism, collision detection (CD) capability is enabled in PBMA protocol. So vehicles which have *Mc* or *Me* CAMs would not initiate backoff process immediately when a collision is detected. Instead they will re-attempt to broadcast in the following time slot. If another collision is detected, vehicles with *Me* CAMs will go through a backoff process and vehicles with *Mc* CAMs will re-attempt one more time in the following slot before going to the backoff process. Such a mechanism is another difference compared to DSRC. The accessing mechanism of our proposed PBMA protocol has been depicted in Fig. 2.

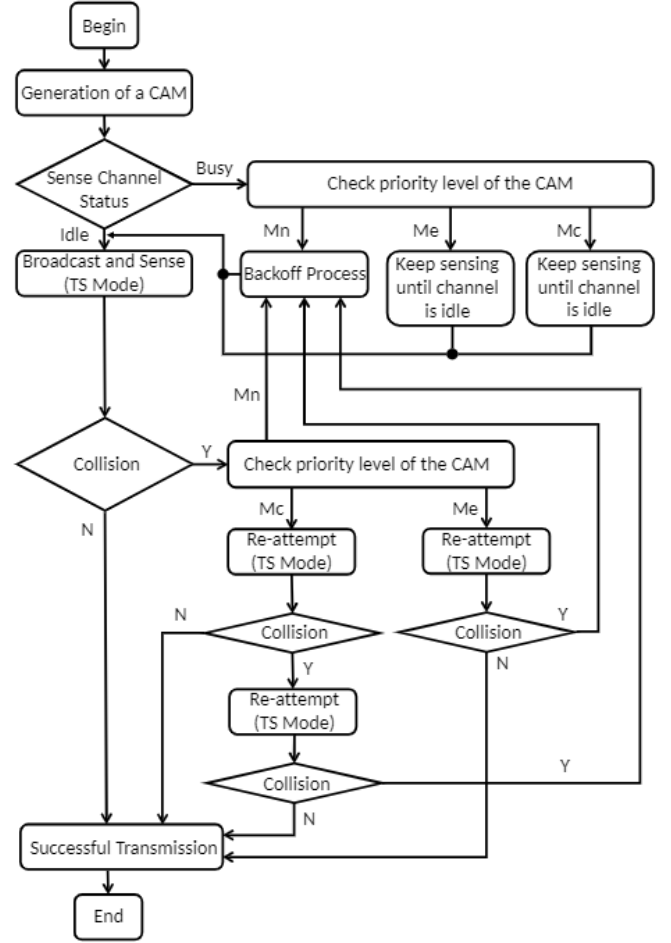

Fig. 2. PBMA prioritised channel access, external collision handling mechanism

The backoff process operates same as DSRC. First, a backoff counter is initialised with an integer random number of time slot which is randomly selected from uniformly distributed CW interval $[0, CW_{max}]$. Vehicles in the backoff process will sense the channel status continuously. If the channel is sensed idle for a slot duration, $t_{slot}$, counter is decremented by 1. Otherwise, counter will be frozen until the channel is sensed idle again for $t_{slot}$ seconds. CAM is not transmitted until the counter reaches zero.

In summary, the proposed FD PBMA protocol has the following differences and advantages compared to DSRC:
- CD capability: FD technology has matured significantly in recent years, and given the less-constrained physical dimensions in OBUs, vehicles in future are expected to be able to detect collisions during transmission by incorporating full-duplex communication.
- Reliability: Vehicles are capable of detecting impending collision of a transmission without ACK messages. They can abort a transmission as soon as a collision is detected, so collision duration would be shortened. In addition, results show our design would work even when SIC is relatively poor.
- Compatibility: Our design does not require ACK messages and further signalling, so it has great compatibility with current DSRC standard.
- Enhanced priority: Our design has conquered the priority challenge between CAMs from different vehicles. In DSRC, priority only exists between messages queued in a vehicle's buffer, when collisions happen between vehicles, all the colliding CAMs are lost, which may lead to severe delay or loss of emergency messages.

## III. SYSTEM MODEL

We consider a VANET in which vehicles broadcast CAMs periodically with a fixed CAM repetition interval $t_{CAM}$. All vehicles are equipped with FD capability. Rayleigh flat fading is assumed to be the channel model between vehicles. The noise component is assumed to be Gaussian, independent and identically distributed (i.i.d.) with zero mean and unit variance. Vehicles are distributed according to a Poisson Point Process (PPP) with density β as shown in Fig. 3. Such an assumption holds when the transmission range of vehicles is larger than the width of the road [12], [26]-[27]. In addition, vehicles have different transmission ($R_{tx}$) and sensing ranges ($R_{sens}$), and sensing range is larger than transmission range. Although the effect of hidden node problem is not eliminated, it can be weakened by increasing the sensing range $R_{sens}$ and setting up thresholds according to the analysis in section IV.

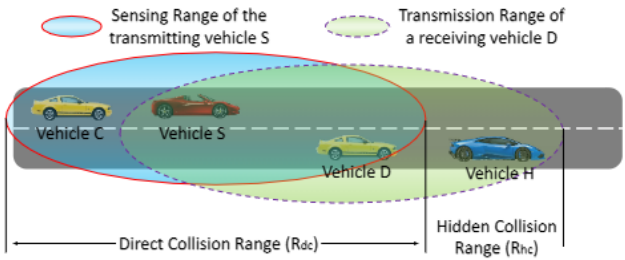

Fig. 3. Demonstration of the VANET model and analysis of sections according to the position of vehicles

From now on when we refer to transmitting vehicle we mean the vehicle which is or going to transmit and sense the channel. Colliding vehicle(s) refer to the vehicle(s) that incur collision due to concurrent transmission. Hidden vehicle(s) are those vehicle(s) that are hidden to the transmitting vehicle (beyond the sensing range of the transmitting vehicle), but their transmissions would cause interference to a generic receiving vehicle.

Fig. 3 demonstrates two different ranges in which transmission of vehicles could possibly collide with the transmission of a generic transmitting vehicle S. Direct collision is defined as the collision of transmission by S with any vehicle (e.g. vehicle C) which is within the sensing range of S. The range of distance in which vehicles may lead to direct collision is called direct collision range and its range is represented by $R_{dc}$. Hidden collision is defined as the collision of transmission by S with any vehicle (e.g. vehicle H) which is beyond the sensing range of S and within the transmission range of a generic receiving vehicle such as D. The range of distance in which vehicles may lead to hidden collision is called hidden collision range and its range is represented by $R_{hc}$. Therefore, the average number of vehicles in the direct collision range is $N_{tx} = 2 \cdot R_{sens} \cdot \beta$ and the average number of vehicles in the hidden collision range is $N_{hc} = (0, d + R_{sens} - R_{tx}) \cdot \beta$.

In our model, sensing is carried out through measuring the energy level of the channel. Since the received energy depends on the distance between the sensing vehicle and the potential concurrent transmitting vehicle(s), to be conservative, we have developed our model to be able to detect the signal collision from the farmost vehicle. Here vehicle S is assumed to be a transmitting and sensing vehicle, whilst vehicles C and D are potentially concurrent transmitting vehicles. It is obvious that S can easily detect the transmission of C if C is also broadcasting as C is close to S. However, D is relatively far away from S. If D is also broadcasting, the detecting of its transmission would be more difficult than detecting that of C. So, we set our thresholds to satisfy the detection of signals sent by D (farmost vehicle). Certain requirements for the received signal-to-interference-plus-noise ratio (SINR), sensing time and SIC capability are found, which will be discussed in section IV. Furthermore, our method performs even better when multiple CAMs are competing for broadcasting at the same time because the energy level of the received signal would be much higher and the colliding signal is much easier to be accurately detected. If the measured energy is less than the threshold $\varepsilon_{th_0}$ which is derived in Section IV, the vehicle knows the channel is free for broadcasting; if the sensed energy is greater than the threshold $\varepsilon_{th_0}$, the vehicle knows there is another vehicle occupying the channel, and will not broadcast until the channel is free. Sensing process continues during the broadcasting period in a FD manner. The measured energy would be compared to an elevated threshold $\varepsilon_{th_1}$ which is dependent on the amount of residual SI after cancellation. If the measured energy is higher than this elevated threshold $\varepsilon_{th_1}$, the vehicle knows its transmission is in collision with another vehicle. Otherwise, the vehicle itself is regarded as the only one using the channel in the network.

However, the aforementioned detection is not perfect. All decisions are made with certain probabilities. Detection probability is defined as the probability that a vehicle successfully detects the presence of an event (an ongoing transmission or a collision) when the event actually takes place, and false alarm probability is defined as the probability that a vehicle falsely declare the presence of an event when the event does not occur. In order to have a high probability of detection, both thresholds (i.e. $\varepsilon_{th_0}$, $\varepsilon_{th_1}$) should be set to a low value. However, this setting will also lead to a high false alarm probability. In other words, we are missing opportunities to transmit. When a collision or false alarm occurs, it is left to the PBMA protocol to decide what course of actions should be followed.

TABLE II.
IMPORTANT NOTATIONS

| Parameters | Notes |
|---|---|
| $N$ | Number of samples |
| $r[n]$ | Received signal at a FD node, where $n = 1, 2, \ldots, N$ |
| $\tau$ | Sensing time |
| $f_s$ | Sampling frequency |
| $w[n]$ | Noise signal with mean zero and variance $\sigma_w^2$ |
| $s_i[n]$ | SI signal with mean zero and variance $\sigma_i^2$ |
| $s[n]$ | transmit signal with mean zero and variance $\sigma_s^2$ |
| $\eta$ | SIC factor |
| $E[.]$ | Expectation operator |
| $\sigma_w^2$ | variance of $w[n]$ ($\sigma_w^2 = E\|w[n]\|^2$) |
| $\sigma_i^2$ | variance of $s_i[n]$ ($\sigma_i^2 = E\|s_i[n]\|^2$) |
| $\sigma_s^2$ | variance of $s[n]$ ($\sigma_s^2 = E\|s[n]\|^2$) |
| $E$ | Energy detection test statistic |
| $\gamma_1$ | measured SNR of the node itself ($\gamma_1 = \frac{\sigma_i^2}{\sigma_w^2}$) |
| $\gamma_2$ | measured SNR from aother node ($\gamma_1 = \frac{\sigma_s^2}{\sigma_w^2}$) |
| $\varepsilon_{th_0}$ | threshold used before transmission |
| $\varepsilon_{th_1}$ | threshold used during transmission |
| $H_i$ | hypothesis $i$ where $i = 0, 1, 2, 3$ |
| $P_{f,bt}$ | probability of false alarm before transmission |
| $P_{f,dt}$ | probability of false alarm during transmission |
| $P_{d,bt}$ | probability of detection before transmission |
| $P_{d,dt}$ | probability of detection during transmission |
| $Q(.)$ | Q function operation |
| $p_i(x)$ | PDF of $E$ under hypothesis $H_i$ |
| $\mu_i$ | mean value of $p_i(x)$ |
| $\sigma_i^2$ | variance of $p_i(x)$ |

In order to make the mathematical formulations clear, we list the important notations in Table. II. Specifically, $\eta$ refers to SIC factor which is the percentage of residual SI power after SIC and it varies between 0 and 1. If $\eta = 0$, it means that SIC is perfect and there is no residual SI.

## IV. MATHEMATICAL ANALYSIS

First of all, we analyse the PHY layer FD simultaneous transmitting and sensing as follows. Four hypotheses for different transmission scenarios are used. Hypothesis $H_0$ is defined as when there is no vehicles broadcasting; $H_1$ is defined as when there is an ongoing transmission from colliding vehicle(s); $H_2$ is defined as when the transmitting vehicle is the only vehicle occupying the channel and $H_3$ is defined as when there are at least 2 vehicles competing for broadcasting.

So the received signal at a FD-enabled vehicle would be

$$r[n] = \begin{cases} w[n]; & H_0 \\ s[n] + w[n]; & H_1 \\ \sqrt{\eta} s_i[n] + w[n]; & H_2 \\ \sqrt{\eta} s_i[n] + s[n] + w[n]; & H_3 \end{cases} \quad (1)$$

The energy detection test statistic is given by

$$E = \frac{1}{N} \cdot \sum_{n=1}^{N} |r[n]|^2. \quad (2)$$

In the sensing phase, Doppler effect would affect detection accuracy. Doppler frequency shift formula is given by

$$\Delta f = \frac{\Delta v}{c} \cdot f_0, \quad (3)$$

where $\Delta f$ denotes the frequency shift, $\Delta v$ is the relative speed between vehicles, c represents the speed of light and $f_0$ refers to the emitted centre frequency.

In addition, required bandwidth for a CAM broadcast can be calculated by Shannon-Hartley theorem:

$$C = B \times \log_2(1 + SNR). \quad (4)$$

Taking the standardised parameters in DSRC [c28] as an example, we assume that the $SNR$ is $1\,dB$, transmit rate is $6\,Mbps$, maximum relative speed between vehicles is $500\,km/h$. Thus, the required bandwidth for a CAM broadcast is $6\,MHz$, and the maximum Doppler frequency shift is approximately $2.731\,kHz$. The results show that only a portion of the allocated $10\,MHz$ bandwidth will be used for a CAM broadcast, and there is enough guard frequency gap between channels.

Therefore, in order to mitigate Doppler effect in the sensing phase, instead of sensing the bandwidth used for broadcast, we increase the sensing bandwidth. The increased sensing bandwidth $B'$, by taking Doppler frequency shift into consideration, is designed to be twice of the maximum Doppler frequency shift, which is given by

$$B' = B + 2 \cdot \Delta f. \quad (5)$$

As shown in Fig. 4, by increasing the sensing window size (i.e. sensing window 2), no information will be lost, as the whole signal falls into the sensing range. On the contrary, if sensing window size remains unchanged (i.e. sensing window 1), due to the effect of Doppler frequency shift, the shifted part of the signal (i.e. shadowed part) goes beyond the sensing range, resulting in losing some information of the signal. In addition, the higher the SNR is, the worse the sensing accuracy will be, as a higher percentage of the signal will go beyond the sensing window, resulting in more energy of the signal cannot be measured. Hence, the measured energy will be lower than the actual energy of the signal. Accordingly, probability of detection will decrease, and probability of false alarm will increase.

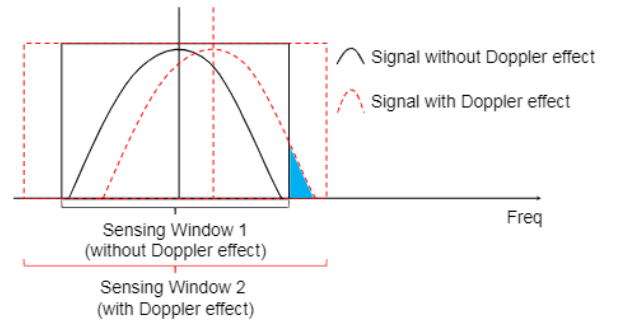

Fig. 4. Demonstration of the increased sensing bandwidth strategy for mitigating Doppler effect on sensing accuracy

Furthermore, this strategy will not be affected by inter-channel interference, because the sensing window size is much smaller than the allocated 10 $MHz$ channel bandwidth.

A comparison of sensing accuracy between two sensing window sizes has been made later in the simulation section.

Then, we analyse probabilities of detection and false alarm before and during transmission with increased sensing window size. Under hypothesis $H_0$, $E$ is a random variable (RV) whose probability density function (PDF) $p_0(x)$ follows a Chi-squared distribution, probability of false alarm can be expressed as [29]

$$P_{f,bt} = Q((\frac{\varepsilon_{th_0}}{\sigma_w^2} - 1) \cdot \sqrt{N}). \quad (6)$$

Under hypothesis $H_1$, probability of detection under this hypothesis is given by

$$P_{d,bt} = Q((\frac{\varepsilon_{th_0}}{\sigma_w^2} - \gamma_2 - 1) \cdot \sqrt{\frac{N}{2\gamma_2+1}}). \quad (7)$$

Under hypothesis $H_2$, similar to $H_1$, probability of false alarm is derived as

$$P_{f,dt} = Q((\frac{\varepsilon_{th_1}}{\sigma_w^2} - \eta^2\gamma_1 - 1) \cdot \sqrt{\frac{N}{2\eta^2\gamma_1+1}}). \quad (8)$$

Under hypothesis $H_3$, similar to the previous hypotheses, probability of detection during transmission is given by

$$P_{d,dt} = Q((\frac{\varepsilon_{th_1}}{\sigma_w^2} - \gamma_2 - \eta^2\gamma_1 - 1) \times \sqrt{\frac{N}{2\eta^2\gamma_1+2\eta^2\gamma_1\gamma_2+2\gamma_2+1}}). \quad (9)$$

Next, we analyse the relationship between thresholds $\varepsilon_{th_0}$ and $\varepsilon_{th_1}$, when increased sensing window size strategy is deployed. Threshold $\varepsilon_{th_0}$ is found from Eq. (7) by calculating the inverse Q function, which is given by

$$\varepsilon_{th_0} = (\frac{Q^{-1}(P_{d,bt})}{\sqrt{\frac{N}{2\gamma_2+1}}} + \gamma_2 + 1) \cdot \sigma_w^2, \quad (10)$$

and threshold $\varepsilon_{th_1}$ is given by Eq. (9) as

$$\varepsilon_{th_1} = (\frac{Q^{-1}(P_{d,dt})}{\sqrt{\frac{N}{2\eta^2\gamma_1+2\eta^2\gamma_1\gamma_2+2\gamma_2+1}}} + \gamma_2 + \eta^2\gamma_1 + 1) \cdot \sigma_w^2. \quad (11)$$

Assume the target probabilities of detection before and during transmission are identical, then the relationship between the two thresholds can be derived as

$$\varepsilon_{th_1} = \frac{\frac{\varepsilon_{th_0}}{\sigma_w^2} - \gamma_2 - 1}{\sqrt{\frac{2\gamma_2+1}{2\eta^2\gamma_1+2\eta^2\gamma_1\gamma_2+2\gamma_2+1}}} + \eta^2\gamma_1 + \gamma_2 + 1. \quad (12)$$

This relationship is depicted in Fig. 5. It shows that the higher the residual SI, the higher the thresholds, and the bigger the difference between the two thresholds would be.

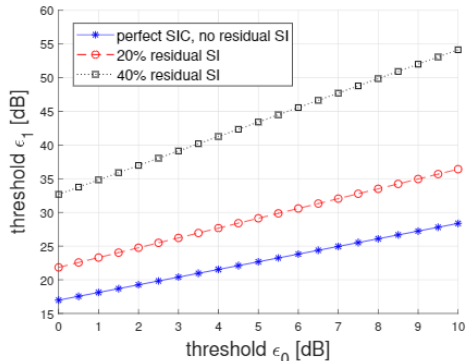

Fig. 5. relationship between threshold $\varepsilon_{th_0}$ and threshold $\varepsilon_{th_1}$

In addition, SIC factor is not always fixed, it may fluctuate due to the imperfection of the hardware or channel variations. For a given SIC factor $\eta_0$ with $\pm m\%$ fluctuation distributed uniformly, with the help of the approximation of the Q function [30], the average probability of false alarm can be calculated by

$$\overline{P_{f,dt}} \approx \frac{1}{2}Q\left(\frac{\varepsilon_{th_1}}{\sigma_w^2} - (\eta_0+m)^2\gamma_1 - 1\right)\sqrt{\frac{N}{2(\eta_0+m)^2\gamma_1+1}} + \frac{1}{2}Q\left(\frac{\varepsilon_{th_1}}{\sigma_w^2} - (\eta_0-m)^2\gamma_1 - 1\right)\sqrt{\frac{N}{2(\eta_0-m)^2\gamma_1+1}}. \quad (13)$$

According to the above analysis and thresholds settings, a vehicle can detect concurrent transmissions with certain probabilities of detection and false alarm. Then the vehicle will schedule its broadcast according to the deployed MAC layer protocol. We formulate collision probability, collision duration, waiting time as well as throughput by deploying three different mechanisms. The first mechanism is the current DSRC standard which uses HD EDCA method. The second method is FD EDCA scheme and the last strategy is our proposed FD PBMA design.

A. Collision Probability

Collisions happen due to direct collisions and hidden collisions. Direct collisions happen in two cases. The first case is when the channel is idle, there are at least two vehicles which have CAMs to broadcast and all of them do not wrongly detect (i.e. no false alarm) the channel status. The other case is when the channel is busy, there is at least one vehicle which has a CAM to transmit, but the vehicle(s) mis-detect(s) the presence of the ongoing transmission. So, the overall direct collision probability is given by the sum of the probabilities in these two cases:

$$P_{dc}(HD) = P_{dc1}(HD) + P_{dc2}(HD). \quad (14)$$

Direct collision probability in the first case (i.e. channel idle) is given by

$$P_{dc1}(HD) = \sum_{i=2}^{i=N_{tx}} P_{idle}(HD)(1-P_f)^i P_s(i), \quad (15)$$

where $P_{idle}(HD)$ represents the probability that the channel is idle. $P_f$ is the false alarm probability, $P_s(i)$ represents the probability that $i$ vehicles broadcast CAMs at the same time.

To find $P_{idle}(HD)$, we first build up a Markov model to evaluate the backoff process, because channel is idle when no vehicle is transmitting, and an arbitrary vehicle is not transmitting in two cases. The first case is where a vehicle has already finished transmission and has nothing to transmit within the current CAM repetition interval. The second case is where a vehicle has something to transmit but it is in the backoff process.

If we treat each backoff process independently, once the counter is decremented to 0, it never goes out of the state. Besides, it is possible to go from any state to state 0. Therefore, the conditions for using an absorbing Markov chain model are satisfied and the model is shown in Fig. 6.

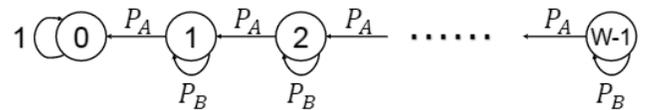

Fig. 6. Markov model for analysing the backoff process

$P_A$ represents the probability to find the channel idle for one slot duration, which is given by

$$P_A = P_{idle}(HD)(1 - P_f) + (1 - P_{idle}(HD))(1 - P_d), \quad (16)$$

where $P_d$ refers to the detection probability.

$P_B$ refers to the probability to find the channel busy for one slot duration, which is given by

$$P_B = P_{idle}(HD)P_f + (1 - P_{idle}(HD))P_d. \quad (17)$$

So the transition matrix $P$ is

$$\begin{bmatrix} P_A & 0 & 0 & \dots & 0 & P_B \\ P_B & P_A & 0 & \dots & 0 & 0 \\ 0 & P_B & P_A & \dots & 0 & 0 \\ \vdots & \vdots & \vdots & \ddots & \vdots & \vdots \\ 0 & 0 & 0 & \dots & P_A & 0 \\ 0 & 0 & 0 & \dots & 0 & 1 \end{bmatrix}.$$

The fundamental matrix $N = (I - Q)^{-1}$ is

$$\begin{bmatrix} 1 - P_A & 0 & 0 & \dots & 0 \\ -P_B & 1 - P_A & 0 & \dots & 0 \\ 0 & -P_B & 1 - P_A & \dots & 0 \\ \vdots & \vdots & \vdots & \ddots & \vdots \\ 0 & 0 & 0 & \dots & 1 - P_A \end{bmatrix}^{-1}.$$

Thus average waiting time for each backoff process is

$$t_{BO} = \frac{1}{W} \cdot \left( \sum_{i=0}^{W-1} \sum_{j=0}^{W-1} (N_{ij}) \right) \cdot t_{slot}, \quad (18)$$

where $i$ and $j$ refer to the row and column index of the fundamental matrix, $W$ refers to the contention window size and $i, j \in [0, W-1]$.

Therefore we get $P_{idle}(HD)$ as

$$P_{idle}(HD) = \sum_{i=0}^{i=N_{tx}} \left( \frac{t_{CAM} - t_{BO} - t_{pkt}}{t_{CAM}} \right)^i \cdot \left( \frac{t_{BO}}{t_{CAM}} \right)^{N_{tx}-i}, \quad (19)$$

where $t_{CAM}$ is the CAM repetition interval and $t_{pkt}$ is transmission duration of a CAM.

Now the only unknown variable is $P_s(i)$. To find $P_s(i)$, we introduce $P_r$ which refers to the probability that there is a CAM waiting to be broadcast at a vehicle, and $P_\sigma$ denoting the probability that the CAM is ready to be broadcast immediately (i.e. backoff counter is zero). Thus $P_s(i)$ is given by

$$P_s(i) = \sum_{i=2}^{i=N_{tx}} 1 - (1 - P_r P_\sigma)^{i-1}. \quad (20)$$

According to the absorbing Markov chain model shown in Fig. 6, $P_\sigma$ is given by

$$P_\sigma = \frac{1}{W^2} \left( \sum_{i=0}^{W-1} \sum_{j=0}^{W-1} (N_{ij}) \right) + \frac{1}{W}, \quad (21)$$

and the last unknown variable $P_r$ can be found from

$$P_r = \frac{1}{t_{CAM}} \Big\{ \frac{W \cdot (1 - P_{idle}(HD))}{2} [(1 - P_s(i))t_{slot} + P_s(i) \\ (t_{slot} + t_{AIFS} + t_{pkt})] + t_{pkt} \Big\}. \quad (22)$$

Similarly we can find the direct collision probability in the second case (i.e. channel busy) as

$$P_{dc2}(HD) = \sum_{i=1}^{i=N_{tx}} (1 - P_{idle}(HD))(1 - P_d)^i P_s(i). \quad (23)$$

Hidden collision probability is approximated in a same way as shown in [23] as double of the transmission time from the vehicles at $R_{dc}$:

$$P_{hc}(HD) = \frac{2(N_{hc}-1)}{t_{CAM}} (t_{AIFS} + t_{pkt})(1 - \frac{P_{dc}(HD)}{2}). \quad (24)$$

Thus the overall collision probability in DSRC is given by

$$P_c(HD) = P_{dc}(HD) + P_{hc}(HD) - P_{dc}(HD)P_{hc}(HD). \quad (25)$$

Then we analyse collision probability ($P_c(FD)$) when FD EDCA is deployed. Compared to HD EDCA mechanism, the only difference is that vehicles can sense the channel whilst broadcasting. So vehicles can abort transmissions and initiate the backoff process as soon as collisions are detected. Direct collision is also composed of the aforementioned two cases (channel idle case and channel busy case), which is given by

$$P_{dc}(FD) = P_{dc1}(FD) + P_{dc2}(FD), \quad (26)$$

where $P_{dc1}(FD)$ represents the collision probability when channel is idle and $P_{dc2}(FD)$ represents the collision probability when channel is busy.

Direct collision probability in the first case (i.e. channel idle) is given by

$$P_{dc1}(FD) = \sum_{i=2}^{i=N_{tx}} P_{idle}(FD)(1 - P_f)^i P_s(i), \quad (27)$$

and direct collision probability in the second case (i.e. channel busy) is given by

$$P_{dc2}(FD) = \sum_{i=1}^{i=N_{tx}} (1 - P_{idle}(FD))(1 - P_d)^i P_s(i). \quad (28)$$

Hidden collision probability is found by the same method as in the analysis of the DSRC standard as

$$P_{hc}(FD) = \frac{2(N_{hc}-1)}{t_{CAM}} (t_{AIFS} + t_{pkt})(1 - \frac{P_{dc}(HD)}{2}). \quad (29)$$

Finally we get the overall collision probability when FD EDCA mechanism is deployed as

$$P_c(FD) = P_{dc}(FD) + P_{hc}(FD) - P_{dc}(FD)P_{hc}(FD). \quad (30)$$

Now we extend our analysis to our proposed PBMA design. Assume normal update, emergency and critical CAMs are generated with probabilities, $P_{gn}$, $P_{ge}$ and $P_{gc}$, respectively. Four cases could lead to direct collisions. The first case is when the channel is busy, there is at least one vehicle which has a CAM to transmit, but the vehicle(s) mis-detect(s) the presence of the ongoing transmission. Direct collision probability in this case is given by

$$P_{dc1}(PBMA) = \sum_{i=1}^{i=N_{tx}} (1 - P_{idle}(PBMA))(1 - P_d)^i P_s(i). \quad (31)$$

The second case is when the channel is idle, there are at least two $Mc$ or $Me$ CAMs generated and at least two of these vehicles do not announce false alarm. Direct collision probability in this case is given by

$$P_{dc2}(PBMA) = \\ \sum_{i=2}^{i=N_{tx}} P_{idle}(PBMA)(P_{gc} + P_{ge})^i (1 - P_f)^i P_s(i). \quad (32)$$

The third case is when the channel is idle, there is one critical or emergency CAM and at least one normal update CAM which is going to transmit at the same time. Direct collision probability in this case is given by

$$P_{dc3}(PBMA) = \\ \sum_{i=1}^{i=N_{tx}} P_{idle}(PBMA)(P_{gc} + P_{gn})^i (1 - P_f)^i P_s(i). \quad (33)$$

The last case is when the channel is idle, there are only at least two normal messages ready to transmit at the same time(no $Mc$ and $Me$). Direct collision is given by

$$P_{dc4}(PBMA) = \\ \sum_{i=1}^{i=N_{tx}} P_{idle}(PBMA) P_{gn}^i (1 - P_f)^i P_s(i). \quad (34)$$

Thus overall direct collision probability is given by

$$P_{dc}(PBMA) = P_{dc1}(PBMA) + P_{dc2}(PBMA) \\ + P_{dc3}(PBMA) + P_{dc4}(PBMA). \quad (35)$$

Similar to the DSRC analysis, we obtain hidden collision probability as

$$P_{hc}(PBMA) = \\ \frac{2}{t_{CAM}} (N_{hc} - 1)(t_{AIFS} + t_{pkt})(1 - \frac{P_{dc}(PBMA)}{2}). \quad (36)$$

Therefore the overall collision probability by deploying the PBMA design is given by
$$P_c(PBMA) = P_{dc}(PBMA) + P_{hc}(PBMA) - P_{dc}(PBMA)P_{hc}(PBMA). \quad (37)$$

### B. Collision Duration

In DSRC whether a direct collision or hidden collision happens, collision lasts for a whole packet time because vehicles are not able to detect collisions. Average collision duration is
$$C_d^A(HD) = P_c(HD) \cdot t_{pkt}. \quad (38)$$

When FD EDCA mechanism is deployed and sensing is considered to be imperfect, since the probability that all vehicles in collision mis-detect for three consecutive time slots is very low, we only consider for up to three consecutive mis-detections. Average collision duration is given by
$$C_d^A(FD) = \\ P_{dc}(FD)t_h(2P_d - P_d^2) + P_{dc}(FD)2t_h(1-P_d)^2(2P_d - P_d^2) + \\ P_{dc}(FD)3t_h(1-P_d)^4(2P_d - P_d^2) + P_{hc}(FD)t_{pkt}. \quad (39)$$

The last strategy is based on our proposed PBMA design. Overall collision duration is given by the sum of collision durations in seven cases. The first case is direct collisions between normal CAMs, the second case is direct collisions between critical and normal CAMs, the third case is direct collisions between emergency and normal CAMs, the fourth case is direct collisions between critical CAMs, the fifth case is direct collisions between emergency CAMs, the sixth case is direct collisions between critical and emergency CAMs, and the last case is hidden collisions.

The first case happens with probability $P_1 = P_{gn}^2$:
$$C_{d,1}^A(PBMA) = \\ P_{dc}(PBMA)[(2P_d - P_d^2)t_h + (1-P_d)^2(2P_d - P_d^2)2t_h \\ + (1-P_d)^4(2P_d - P_d^2)3t_h]. \quad (40)$$

The second case happens with probability $P_2 = 2P_{gc}P_{gn}$:
$$C_{d,2}^A(PBMA) = \\ P_{dc}(PBMA)[P_d t_h + (P_d^3 - P_d^2 + P_d) \cdot 2t_h + \\ (-2P_d^6 + 9P_d^5 - 15P_d^4 + 13P_d^3 - 7P_d^2 + 2P_d) \cdot 3t_h. \quad (41)$$

The third case happens with probability $P_3 = 2P_{ge}P_{gn}$:
$$C_{d,3}^A(PBMA) = \\ P_{dc}(PBMA)[P_d t_h + (P_d^3 - P_d^2 + P_d) \cdot 2t_h + \\ (P_d^5 - 5P_d^4 + 9P_d^3 - 7P_d^2 + 2P_d) \cdot 3t_h. \quad (42)$$

The fourth case happens with probability $P_4 = P_{gc}^2$:
$$C_{d,4}^A(PBMA) = \\ P_{dc}(PBMA)[(2P_d - P_d^2)3t_h + (2P_d - P_d^4 + 4P_d^3 - 5P_d^2) \\ 4t_h + (-P_d^6 + 6P_d^5 - 14P_d^4 + 16P_d^3 - 9P_d^2 + 2P_d)5t_h]. \quad (43)$$

The fifth case happens with probability $P_5 = P_{ge}^2$:
$$C_{d,5}^A(PBMA) = \\ P_{dc}(PBMA)[(2P_d - P_d^2)2t_h + (2P_d - P_d^4 + 4P_d^3 - 5P_d^2) \\ 3t_h + (-P_d^6 + 6P_d^5 - 14P_d^4 + 16P_d^3 - 9P_d^2 + 2P_d)4t_h]. \quad (44)$$

The sixth case happens with probability $P_6 = 2P_{gc}P_{ge}$:
$$C_{d,6}^A(PBMA) = \\ P_{dc}(PBMA)[P_d \cdot 2t_h + (-P_d^2 + P_d) \cdot 3t_h + \\ (-P_d^4 + 4P_d^3 - 5P_d^2 + 2P_d) \cdot 4t_h]. \quad (45)$$

The last case happens with probability $P_{hc}(PBMA)$:
$$C_{d,7}^A(PBMA) = t_{pkt}. \quad (46)$$

Finally we find the average collision duration as:
$$C_d^A(PBMA) = \sum_{i=1}^{7} P_{idle} \cdot C_{d,i}^A(PBMA). \quad (47)$$

### C. Waiting Time

We define average waiting time as the time duration a packet stays in the buffer, which includes the sensing delay and backoff time. Given the average waiting time for each backoff process in Eq. (18), the overall average waiting time in DSRC can be calculated as
$$T_w(HD) = P_{idle}(HD)t_h + (1 - P_{idle}(HD))(t_h + t_{BO}). \quad (48)$$

In the FD EDCA strategy, average waiting time is attributed to four different cases. The first case is when the channel is idle and no false alarm happens. So the waiting time would be one sensing duration:
$$T_{w,1}(FD) = P_{idle}(FD) \cdot (1 - P_f) \cdot t_h. \quad (49)$$

The second case is when the channel is sensed as busy whilst it is actually busy:
$$T_{w,2}(FD) = (1 - P_{idle}(FD))P_d(t_h + t_{BO}) + \\ (1 - P_{idle}(FD))^2 P_d^2 (2t_h + t_{BO} + t'_{BO}) + \cdots, \quad (50)$$

where $t'_{BO}$ refers to the second continuous backoff duration.

The third case is when false alarm happens:
$$T_{w,3}(FD) = P_{idle}(FD)P_f(t_h + t_{BO}) + \\ P_{idle}(FD)P_f(1 - P_{idle}(FD))P_d(2t_h + t_{BO} + t'_{BO}) + \cdots. \quad (51)$$

The last case is when mis-detection happens, so the vehicle starts transmission but it detects the collision during its transmission:
$$T_{w,4}(FD) = (1 - P_{idle}(FD))^2 P_m P_d(2t_h + t_{BO}) \\ + (1 - P_{idle}(FD))^3 P_m P_d^2 (3t_h + t_{BO} + t'_{BO}) + \cdots. \quad (52)$$

Finally we find the overall average waiting time as the sum of all above four cases:
$$T_w(FD) = \sum_{i=1}^{4} T_{w,i}(FD). \quad (53)$$

Now we analyse the average waiting time of the PBMA mechanism. When a normal update message is generated, the waiting time is equal to the waiting time by using the FD EDCA method:
$$T_{w,1}(PBMA) = P_{gn} \cdot T_w(FD). \quad (54)$$

When a critical message is generated and the channel is sensed as busy, the transmitting vehicle would keep sensing and broadcast as soon as the current transmission is finished. So the average waiting time in this case would be half the packet transmission duration. Otherwise when the channel is sensed as idle, the waiting time would be one sensing duration. In addition, because sensing is not perfect, false alarm and mis-detection could happen. In the false alarm case, vehicles which have critical CAMs will not go to the backoff process. The average waiting time would be half the packet transmission duration. In the mis-detection case, since the probability that continuous mis-detection occurs is small, we assume a correct detection with probability $P_d$ in the first transmission and sensing slot. Then the vehicle would go to the backoff process, and the average waiting time in this case would be two sensing durations plus the backoff time. Therefore, the overall waiting time is given by
$$T_{w,2}(PBMA) = P_{gc}(1 - P_{idle}(PBMA))P_d \frac{t_{pkt}}{2} +$$

$$P_{gc}P_{idle}(PBMA)P_d t_h +$$
$$P_{gc}P_{idle}(PBMA)P_f \frac{t_{pkt}}{2} +$$
$$P_{gc}P_{idle}(PBMA)^4 P_m P_d^3 (4t_h + t_{BO}) +$$
$$P_{gc}P_{idle}(PBMA)^5 P_m P_d^4 (5t_h + t_{BO} + t'_{BO}) + \cdots. \quad (55)$$

When the generated CAM is $Me$, average waiting time would be

$$T_{w,3}(PBMA) =$$
$$P_{ge}(1 - P_{idle}(PBMA))P_d \frac{t_{pkt}}{2} +$$
$$P_{ge}P_{idle}(PBMA)P_d t_h +$$
$$P_{ge}P_{idle}(PBMA)P_f \frac{t_{pkt}}{2} +$$
$$P_{ge}P_{idle}(PBMA)^3 P_m P_d^2 (3t_h + t_{BO}) +$$
$$P_{ge}P_{idle}(PBMA)^4 P_m P_d^3 (4t_h + t_{BO} + t'_{BO}) + \cdots. \quad (56)$$

Therefore, the overall average waiting time is given by the sum of the above three time durations:

$$T_w(PBMA) = \sum_{i=1}^{3} T_{w,i}(PBMA). \quad (57)$$

### D. Throughput

We define the system throughput as the total number of successful broadcast packets within a CAM repetition interval, which can be calculated as

$$SR = N_{tx} \cdot \left(\frac{t_{CAM} - T_w}{t_{CAM}}\right) \cdot (1 - P_c). \quad (58)$$

### E. Evaluation of $Mc$ Transmission

In addition to average system performance, it is also important to compare the performance of transmitting critical messages in different mechanisms. Since the successful delivery of $Mc$ within a short amount of time is the key to avoid accidents.

The formulations of transmitting $Mc$ in the legacy HD EDCA and FD EDCA mechanisms are the same as shown in the previous section, because there is no priority between CAMs from different vehicles. However, our proposed PBMA mechanism operates differently, for which the analysis is shown as follows.

1) *Collision Probability:* When $Mc$ is generated, collision happens in three cases. The first case is mis-detection. The second case is simultaneous start of transmission when channel is idle. The last case is due to hidden collision. Therefore the collision probability of critical messages is given by

$$P_c(M_c) = P_{gc}(1 - P_{idle}(PBMA))(1 - P_d) + P_{gc}P_{idle}(PBMA)P_d P_s(i) + P_{hc}(PBMA). \quad (59)$$

2) *Collision Duration:* Critical CAMs could collide with normal, critical and emergency CAMs, and collision duration for each case is different. The corresponding collision durations are given by Eq. (41), Eq. (43) and Eq. (45), respectively. Besides, collision duration due to hidden collision is shown in Eq. (46). Therefore the average collision duration of critical messages is given by

$$C_d^A(M_c) = \sum_{i=2,4,6,7} P_i \cdot C_d^A(PBMA). \quad (60)$$

3) *Waiting Time:* Average waiting time of $Mc$ has already been analysed and its formulation is shown in Eq. (55).

## V. SIMULATION RESULTS

Following the mathematical analysis, we now evaluate our proposed method through simulations. Vehicles are generated and movements are simulated in SUMO, data is then imported into MATLAB. The PHY and MAC layers of DSRC and our proposed PBMA protocol are simulated in MATLAB. Simulation parameters are shown in Table III.

TABLE III
SIMULATION PARAMETERS

| Parameters | Values |
|---|---|
| Target $P_{d,bt}$ & $P_{d,dt}$ | 90% |
| Modulation scheme | BPSK, QPSK |
| SNR1 | $+ 10\ dB$ |
| SNR2 | $(-20) – 0\ dB$ |
| Residual SI | 0% – 40% |
| Vehicle density | $0 – 200\ vehicles/km$ |
| Relative speed | $0 – 500\ km/h$ |
| Transmission rate | $6\ Mbps$ |
| CAM length | $350\ bytes/pkt$ |
| Arbitrary Inter-Frame Space | $58\ \mu sec$ |
| Slot duration | $13\ \mu sec$ |
| CAM repetition interval | $100\ msec$ |

Fig. 7 shows a good match between the simulated and theoretical values, which verifies our mathematical analysis to be correct and accurate. Furthermore, when residual SI becomes stronger, in order to achieve the same detection probability, thresholds are set to higher values because more energy (from SI signal) is received. In addition, we can see that a small variation of the threshold would result in a huge deviation in the probabilities even when SIC is perfect. For example, if the threshold changes from 1 $dB$ to 1.025 $dB$, such a small change would lead to a 45% drift for the probabilities of detection and false alarm. This result highlights that the calculation of the thresholds is of great importance and should be done as accurately as possible.

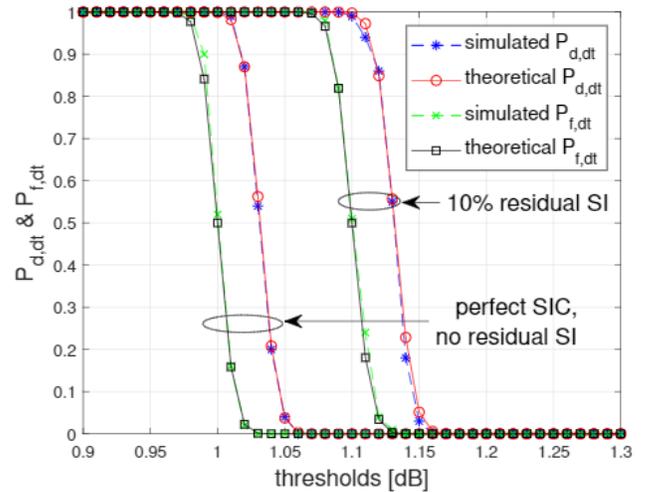

Fig. 7. Probabilities of detection $P_{d,dt}$ and false alarm $P_{f,dt}$ vs. threshold $\varepsilon_{th_1}$ under different SIC assumptions

Fig. 8 shows the impact of mobility on the detection probability. First of all, if vehicles are equipped with dynamic threshold and increased sensing window size, target detection probability (i.e. 90%) can be met. However, if the size of the sensing window remains unchanged, detection probability drops as the relative speed of vehicles increases, since a higher percentage of energy of the signal goes beyond the sensing window, and hence cannot be measured due to the Doppler frequency shift effect.

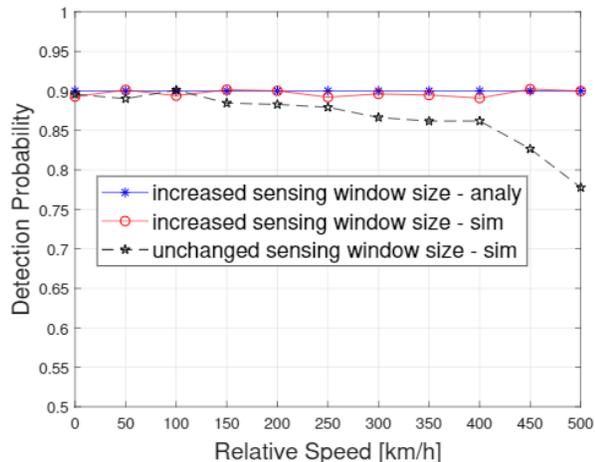

Fig. 8. Impact of mobility on the detection probability

Fig. 9 and Fig. 10 illustrate the significant impact of transmit power and the difference between two threshold setting strategies. One strategy is based on our proposed method where thresholds are dynamically changing, whilst the other strategy corresponds to the fixed threshold method. For the fixed threshold strategy, along with the rise of the transmit power, detection probability increases at the cost of having a high and unacceptable false alarm probability. Our proposed method has a lower detection probability which is still in the acceptable range. But because the threshold is also increasing with the rise of the measured SNR, false alarm probability would decrease at the same time. To summarise, although our proposed method will sacrifice some detection probability by dynamically changing the threshold, a much better false alarm probability would be rewarded, whilst keeping the probability of detection in an acceptable range.

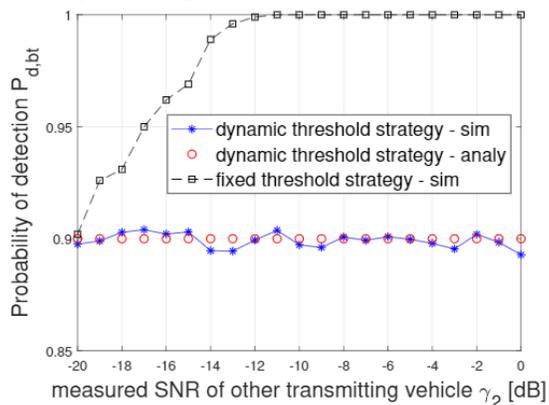

Fig. 9. $P_{d,bt}$ vs. measured SNR before transmission

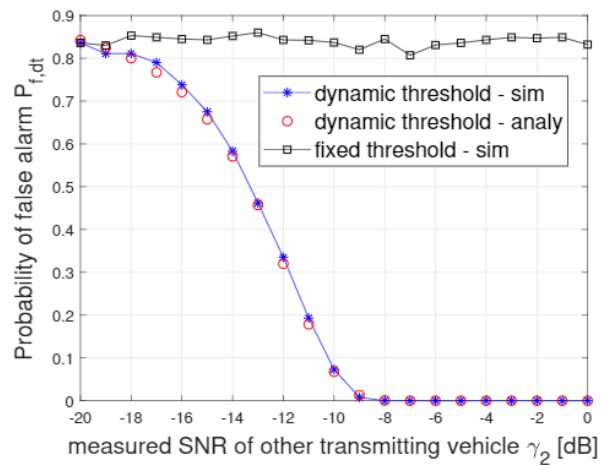

Fig. 10. $P_{f,dt}$ vs. measured SNR during transmission

Fig. 11 shows the effect of residual SI on probabilities of detection and false alarm. Firstly, target detection probability is achievable regardless of SIC by dynamically changing the threshold. However, when η increases, false alarm probability increases too since more energy is received. In order to achieve detection probability to be at least 90% and false alarm probability to be at most 10%, our model would have acceptable performance when SIC is less than 15%. In other words, our system does not operate only when SIC is extremely well, it also works when SIC is relatively poor.

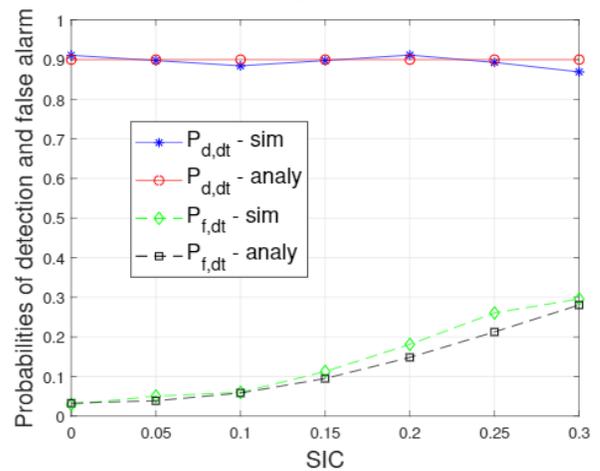

Fig. 11. Probabilities of detection $P_{d,dt}$ and false alarm $P_{f,dt}$ vs. SIC factor η

Fig. 12 highlights the negative effect of SIC fluctuation. When 10% random SIC fluctuation is considered, both probabilities of detection and false alarm become worse. Thus we should carefully consider SIC fluctuation when deploying the scheme.

Fig. 13 shows the impact of the sensing time on the precision of detection. By setting the thresholds properly, system can achieve the target detection performance. Meanwhile, the longer the sensing time is, the lower will be the chance for the system to wrongly alarm an impending collision. This is because we are measuring and averaging the received energy over a longer period of time, which gives a more accurate result. Another way to reduce the false alarm probability is to increase the sampling frequency, since the number of samples is equal to the production of sensing time and sampling frequency ($N = \tau \cdot f_s$). However, the accuracy cannot be improved by only increasing $f_s$. When the number of samples taken is large enough, more samples would not add to accuracy of the measured energy level.

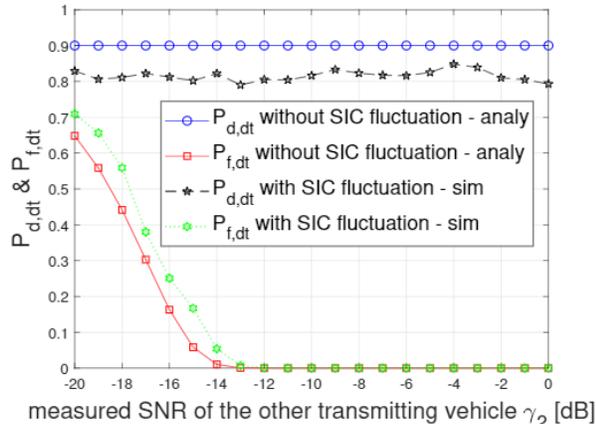

Fig. 12. Probabilities of detection $P_{d,dt}$ and false alarm $P_{f,dt}$ vs. measured SNR during transmission with 10% SIC fluctuation

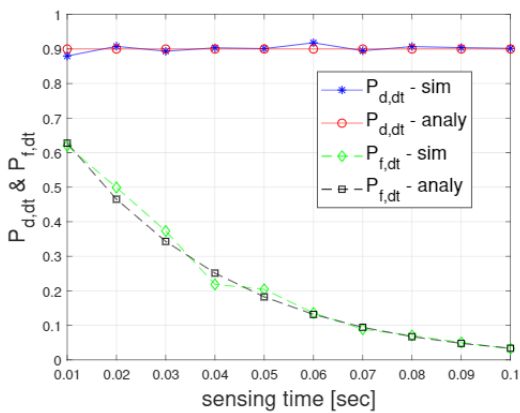

Fig. 13. Probabilities of detection $P_{d,dt}$ and false alarm $P_{f,dt}$ vs. sensing time during transmission

Fig. 14 to Fig. 17 respectively show the average collision probability, collision duration, waiting time and system throughput against the vehicle density by deploying different mechanisms. As shown in Fig. 14, when vehicle density increases, collision probability will go up accordingly due to the fact that more vehicles are competing for broadcasting at the same time. In addition, when a collision happens, it lasts longer (i.e. collision duration is longer) and the waiting time is also longer compared to the network where fewer vehicles exist, as shown in Fig. 15 and Fig. 16.

Furthermore, it can be seen that FD EDCA and our proposed PBMA mechanism outperform DSRC in terms of collision probability and collision duration. This is because CD capability is not enabled in DSRC and collisions cannot be detected through FD technology. DSRC provides the best performance in terms of waiting time as shown in Fig. 16, but a large portion of transmissions are in collisions, and CAMs are lost. This conclusion is drawn from Fig. 14 and Fig. 17, where DSRC gives the worst performance in terms of collision probability and successful packet delivery rate. In other words, compared to FD EDCA and FD PBMA methods, DSRC can broadcast messages quicker with more collisions. Comparing FD PBMA to FD EDCA method, FD PBMA has a slightly higher collision probability and collision duration, which is due to the fact that vehicles which have $M_c$ or $M_e$ would re-attempt to broadcast immediately in FD PBMA method. But FD PBMA has a shorter average waiting time as shown in Fig. 15. The system throughput in Fig. 17 shows the overall performance of different mechanisms, in which PBMA method provides the highest throughput.

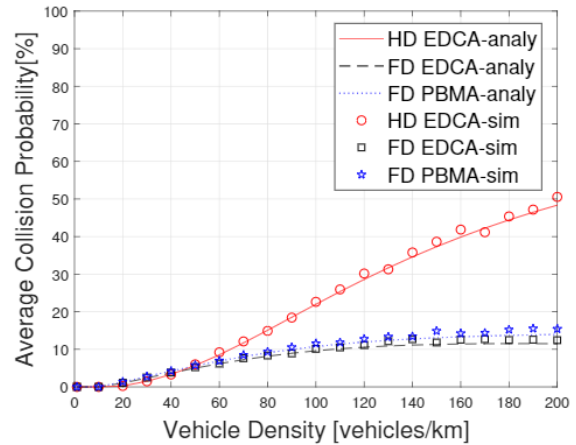

Fig. 14. Average collision probability vs. vehicle density

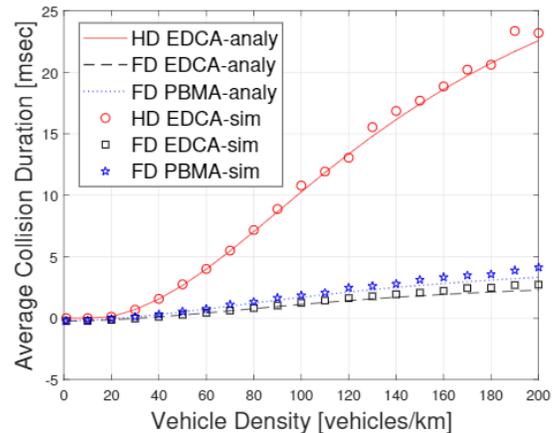

Fig. 15. Average collision duration vs. vehicle density

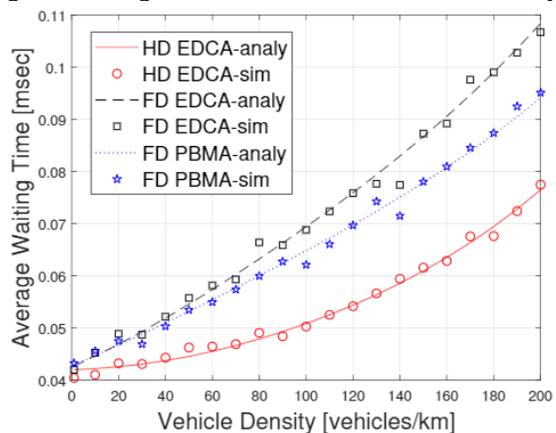

Fig. 16. Average waiting time vs. vehicle density

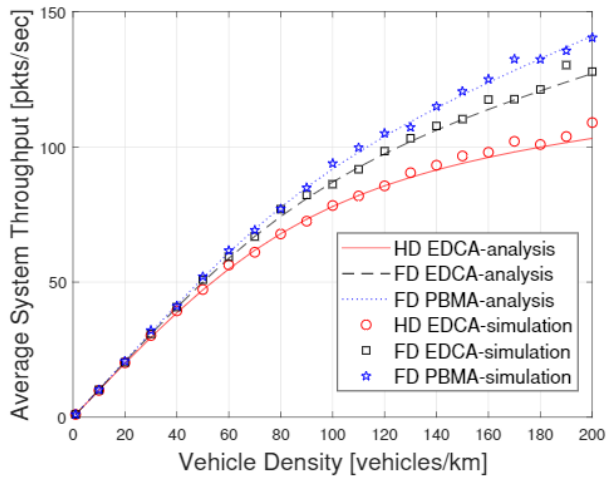

Fig. 17. Average system throughput vs. vehicle density

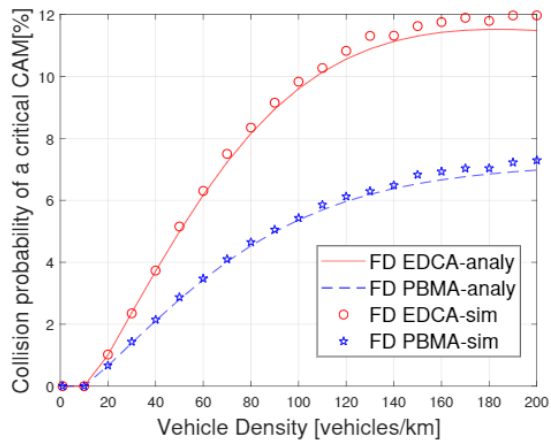

Fig. 18. Collision probability of a *Mc* vs. vehicle density

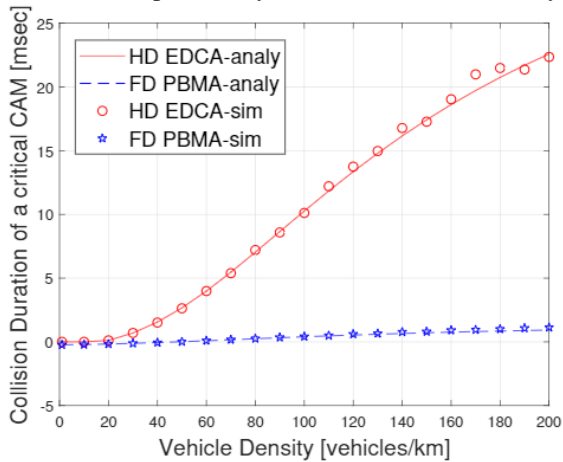

Fig. 19. Collision duration of a *Mc* vs. vehicle density

In addition to average system performance, it is even more important to observe the delivery performance of critical messages, because critical messages could be the last broadcast warning message before potential accidents. Successful delivery of critical messages within a short amount of time will lead to a totally different result. Collision probability, collision duration as well as the waiting time of broadcasting a critical CAM are demonstrated in Fig. 18 to Fig. 20. We did not plot the performance of broadcasting critical messages by deploying DSRC because FD EDCA method is already shown to outperform DSRC.

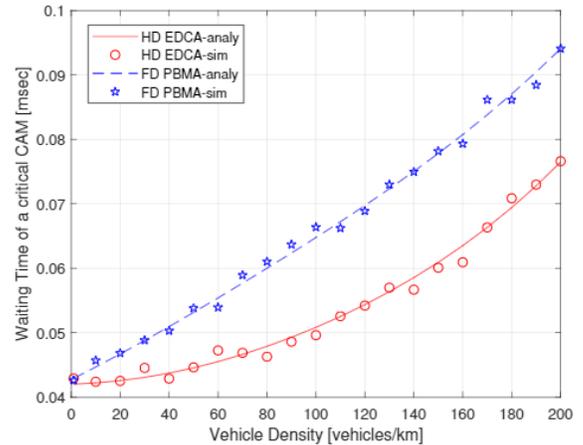

Fig. 20. Waiting time of a *Mc* vs. vehicle density

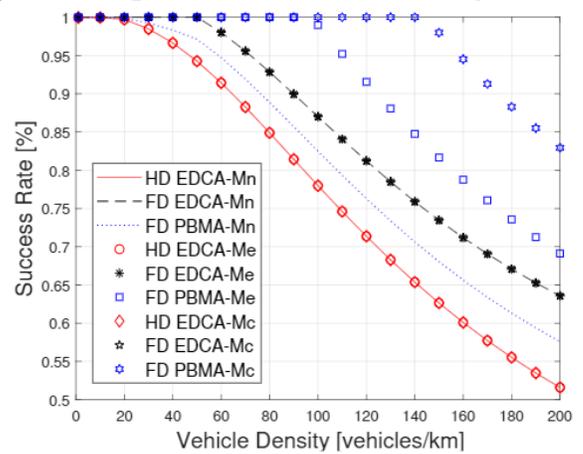

Fig. 21. Success rates of transmitting different types of messages in different mechanisms

By analysing Fig. 18 to Fig. 20, we can see that FD PBMA has significantly improved the delivery of critical CAMs compared to FD EDCA, because FD PBMA can broadcast critical messages with a much smaller collision probability, a much shorter collision duration and a much shorter waiting time too. In addition, it is shown in Fig. 21 that the success rates of transmitting critical messages as well as emergency messages by deploying our proposed FD PBMA design are both significantly enhanced and are higher than those in DSRC and FD EDCA methods. However, the successful delivery rate of broadcasting normal update messages by deploying FD PBMA is lower than that in FD EDCA method and higher than that in the DSRC standard. In other words, when there is a vehicle status change, no matter the change is sudden (critical message) or gentle (emergency message), FD PBMA provides the best performance on letting other vehicles realise such a change from normal cruising activity.

## VI. CONCLUSION

In this paper we proposed a cross-layer design for future V2X networks. By deploying FD technology and setting thresholds as well as sensing bandwidth according to our design, a vehicle can detect and avoid collisions without losing too many opportunities to transmit useful data. Two thresholds which are dynamically changing and an increased sensing window size have been formulated for detection of channel status and collisions. Furthermore, a novel prioritisation scheme dedicated for CAMs and a novel MAC protocol named PBMA are proposed to schedule the access according to detection results and priorities of messages. Comparisons between DSRC, FD EDCA and our proposed PBMA design have been made through both mathematics and simulations in terms of average collision probability, collision duration, waiting time and throughput. Especially, the delivery of critical messages before a potential accident has also been analysed thoroughly. Results have shown that our design works well even when SIC is poor. Meanwhile, our PBMA design also has an overall better performance: The delivery of CAMs before a potential accident has been enhanced and accidents can be further avoided.

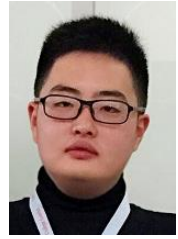

**Junwei Zang** received the B.Eng. degree in Electronic Engineering from University of Leeds, Leeds, U.K., in 2016 and the M.Sc. degree in Telecommunications and Internet Technology from King's College London, U.K., in 2017. He is currently working towards the Ph.D. degree in Telecommunications with the Department of Engineering, Centre for Telecommunications Research, in King's College London, London, U.K., where he has been strongly involved in the EPSRC SENSE project. His current research interests include vehicular communications and full-duplex wireless communications.

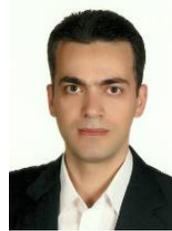

**Vahid Towhidlou** received the B.Sc. degree in Electrical Engineering from the Khajeh Nasir University of Technology, Tehran, Iran, in 1997, the M.Sc. degree in Telecommunication Engineering from Ferdowsi University, Mashhad, Iran, in 2000, and the Ph.D. degree in Telecommunication Engineering from King's College London, U.K., in 2016, where he has been a Research Associate with the Department of Informatics, Centre for Telecommunication Research since 2016. His current research interests lie in the area of cognitive radio, communication for vehicular networks, and full-duplex wireless communication.

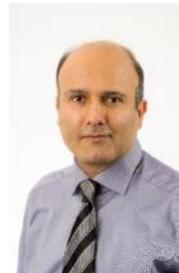

**Mohammad R. Shikh-Bahaei** received the B.Sc. degree from the University of Tehran, Tehran, Iran, in 1992, the M.Sc. degree from the Sharif University of Technology, Tehran, in 1994, and the Ph.D. degree from King's College London, U.K., in 2000. He has worked for two start-up companies, and National Semiconductor Corporation, Santa Clara, CA, USA (currently part of Texas Instruments, Inc.), on the design of third-generation mobile handsets, for which he has been awarded three U.S. patents as an inventor and a co-inventor, respectively. He joined King's College London as a Lecturer in 2002, where he is currently a Full Professor. He has been engaged in research in the area of wireless communications and signal processing for 25 years both in academic and industrial organisations. His research interests include full-duplex and cognitive dense networking, visual data communications over the Internet of Things, applications in healthcare, and secure communication over autonomous vehicle/drone networks. He is a Founder and the Chair of the Wireless Advanced (formerly, SPWC) International Conference from 2003 to 2018. He is a fellow of IET.